\documentclass[
    twocolumn, 
    tighten, 
    times, 
    twocolappendix
    ]{aastex631}

\usepackage{subfigure}
\usepackage[dvipsnames, svgnames]{xcolor}
\usepackage{lipsum}
\usepackage{soul}
\usepackage{amsmath}

\sethlcolor{yellow!50}

\shorttitle{$\gamma$-ray and $\nu$ emission in the Galactic plane}
\shortauthors{Das et al.}
\graphicspath{{./}{figures/}}

\begin{document}

\vspace*{-1.0cm}

\title{TeV--PeV Gamma-ray and Neutrino Emission in the Galactic Plane}

\author[0000-0001-5796-225X]{Saikat Das}
\affiliation{Department of Physics, University of Florida, Gainesville, FL 32611, USA}
\email{saikatdas@ufl.edu}

\author[0000-0002-1188-7503]{Nayantara Gupta}
\affiliation{Astronomy \& Astrophysics Group, Raman Research Institute, Sadashivanagar, Bangalore 560080, Karnataka, India}
\email{nayan@rri.res.in}

\author[0000-0002-0458-7828]{Siyao Xu}
\affiliation{Department of Physics, University of Florida, Gainesville, FL 32611, USA}
\email{xusiyao@ufl.edu}




\begin{abstract}
We model the LHAASO observation of diffuse TeV--PeV $\gamma$ rays in the Galactic plane as the sum of unresolved leptonic emission from pulsar wind nebulae and hadronic emission from supernova-injected cosmic-ray (CR) protons. 
We investigate uncertainties in the radial distribution of the infrared component of the interstellar radiation field (ISRF), 
using profiles with enhanced photon densities in the inner Galaxy. We quantify their effects on $\gamma\gamma$ attenuation of the diffuse $\gamma$-ray emission. 
The alternative ISRF models affect the LHAASO diffuse fit only modestly, as the analysis excludes the Galactic center direction and applies source masks in the Galactic plane.  Using the hadronic normalization inferred from the LHAASO fit for various ISRF models, the associated $pp$ neutrino emission remains consistent with the IceCube all-sky measurement, while the flux from the Galactic Ridge region remains compatible with current ANTARES and KM3NeT constraints. Since the modified infrared profiles differ most strongly toward the inner Galaxy, we also examine their impact on inverse-Compton emission from point sources near the central molecular zone. These same models can noticeably modify the hadronic and inverse-Compton $\gamma$-ray emission above $\sim\!10$ TeV from sources in the central region. Future KM3NeT observations, combined with $\gamma$-ray measurements of individual sources, can probe the inner-Galaxy CR population and constrain the radial distribution of the ISRF near the Galactic Center.
\end{abstract}



\section{Introduction} \label{sec:intro}

The diffuse $\gamma$-ray emission of the Milky Way has long served as a calorimeter of Galactic cosmic rays (CRs), encoding the imprint of their interactions with gas and radiation fields \citep[see, e.g.,][]{Blasi:2013rva, BeckerTjus:2020xzg, Gabici:2019jvz}. Leptonic emission from CR electrons includes bremsstrahlung in the interstellar gas at GeV energies and inverse-Compton (IC) scattering of starlight, infrared (IR), and cosmic microwave background (CMB) photons at $\gamma$-ray energies up to several hundred TeV. Above a few tens of GeV, the contribution of hadronuclear interactions ($pp$) becomes increasingly important through $\pi^0$-decay $\gamma$ rays. A diffuse multi-TeV excess reported by Milagro in the Cygnus region suggests either a hard local CR spectrum or a contribution from unresolved sources \citep{Abdo:2008if}. Fermi-LAT observations revealed an inner-Galaxy excess above a few GeV relative to GALPROP-based diffuse-emission models \citep{2012ApJ...750....3A}. ARGO-YBJ measurements are broadly consistent with the Fermi-LAT diffuse emission expectation, and do not show a sub-TeV excess outside the Cygnus region \citep{ARGO-YBJ:2015cpa}. H.E.S.S. has also detected large-scale TeV emission along the inner Galactic plane, interpreted as a combination of diffuse hadronic emission and unresolved sources \citep{HESS:2014ree}.

The first detection of a diffuse sub-PeV $\gamma$-ray component from the Galactic disk was reported by Tibet AS$\gamma$, pointing to Galactic CR PeVatrons \citep{TibetASgamma:2021tpz}. The plausibility of a hadronic origin motivated searches for a Galactic component in the IceCube TeV--PeV neutrino data. \citet{Razzaque:2013uoa} associated 5 out of 21 showerlike IceCube events with the Galactic Center region, while \citet{Joshi:2013aua} found that Galactic CR interactions with matter can account for at most $0.1$ of the observed IceCube events. Using diffuse TeV--PeV $\gamma$-ray limits, \citet{Ahlers:2013xia} argued that the IceCube excess is mostly extragalactic, although subdominant Galactic contributions remain possible. In light of the Tibet AS$\gamma$ measurement, it was pointed out that there is a possible tension with the nondetection of Galactic neutrinos by IceCube, which may be alleviated by an extra hard CR component associated with Cygnus Cocoon or other Galactic PeV sources \citep{Liu:2021lxk}. The Galactic-plane contribution to the all-sky neutrino flux
associated with the Tibet AS$\gamma$ diffuse emission has been shown
to be $\lesssim 5$--$10\%$ around $100$ TeV
\citep{Fang:2021ylv}.

Spatially dependent CR transport has been invoked to explain the hard inner-Galaxy $\gamma$-ray flux and to predict an enhanced Galactic neutrino component \citep{Gaggero:2015xza}. Subsequent studies examined the dependence on the radial CR density profile and inner-Galaxy spectral hardening \citep{Pagliaroli:2016lgg, Lipari:2018gzn}. Unresolved sources and the diffuse interstellar component can also account for the Tibet AS$\gamma$ data without requiring a hardening of the CR spectrum toward the Galactic center \citep{Steppa:2020qwe, Vecchiotti:2021yzk}.
More recently, LHAASO has observed diffuse $\gamma$-ray emission extending to PeV energies from both the inner and outer Galactic plane,
with a spectral break at $\sim30$ TeV \citep{LHAASO:2023gne, LHAASO:2024lnz}.

IceCube has reported evidence for diffuse neutrino emission from the Galactic plane \citep{IceCube:2023ame}. A joint HAWC--IceCube analysis found no significant neutrino counterparts to 22 Galactic $\gamma$-ray sources and constrained the hadronic fraction of five sources \citep{HAWC:2024kkc}.
Neutrino observations of LHAASO sources constrain their hadronic
contribution \citep{Huang:2021hjc}. Additional unresolved Galactic source components have been introduced in multimessenger modeling of the LHAASO diffuse emission and the IceCube Galactic-plane neutrino signal \citep{Shao:2023aoi}. \citet{Yan:2023hpt} argued for a significant non-hadronic contribution, plausibly from pulsar halos, whereas data-driven estimates suggest that unresolved pulsar sources cannot dominate the LHAASO diffuse emission above a few tens of TeV \citep{Kaci:2024lwx}. Multimessenger arguments suggest that the Galactic-plane neutrino signal is compatible with a hadronic diffuse $\gamma$-ray component, while resolved $\gamma$-ray sources alone are unlikely to account for the IceCube flux \citep{Fang:2023ffx,Fang:2023azx}.

Population-based arguments suggest that the sources of the Galactic hadronic CR flux may not be the brightest TeV--PeV $\gamma$-ray sources \citep{Sudoh:2022sdk}.
The diffuse flux may instead arise from CR propagation in the interstellar medium (ISM) followed by interactions with gas \citep{Lipari:2024pzo}, or from CR trapping in source cocoons \citep{Ambrosone:2025wxc}. Recent work also suggests that the Galactic $\gamma$-ray and neutrino emission from GeV to PeV energies can be explained largely by the Galactic CR sea, without a dominant additional source population \citep{Luque:2022buq, DeLaTorreLuque:2025zsv}. These different interpretations leave the origin of the Galactic TeV--PeV emission unsettled, further motivating careful treatment of propagation effects that modify the observed $\gamma$-ray spectrum.

In this work, we quantify how uncertainties in the radial distribution of the IR component of the interstellar radiation field (ISRF) affect the observed $\gamma$-ray spectrum from the Galactic plane. We model the 
LHAASO diffuse TeV--PeV $\gamma$-ray spectrum as the sum of a leptonic contribution from unresolved pulsar wind nebulae (PWNe) 
and a hadronic contribution from supernova-injected CR protons diffusing through the Galaxy. Using a recent gas-density reconstruction, we calculate the resulting $\gamma$-ray emission and associated $pp$ neutrino flux. 
Motivated by the possibility of an enhanced 
ISRF in the inner Galaxy
\citep{Misiriotis:2006qq,Vernetto:2016alq,Gupta:2024hxl}, we 
consider IR profiles with enhanced photon densities in the inner Galaxy. We evaluate their impact
on the $\gamma\gamma$ optical depth and hence on the fit to the observed diffuse $\gamma$-ray spectrum. The total optical depth is the sum of that due to starlight, IR, and CMB.

Since these IR profiles differ most strongly toward the Galactic Center, we also examine the effect of an enhanced ISRF on
IC emission from point sources in the inner few hundred parsecs, near the central molecular zone (CMZ). 
We further use the hadronic component of the combined leptonic and hadronic fit to the LHAASO diffuse emission to estimate the neutrino flux from the Galactic Ridge ($|\ell|<30^\circ$, $|b|<2^\circ$). This region is a key target for upcoming imaging atmospheric Cherenkov detectors such as CTA and also KM3NeT, which has a favorable view of the Galactic
Center with upgoing muon tracks. Future KM3NeT observations will test the unresolved-source contribution, particularly toward the Galactic
Center.

We outline the model considerations in Sec.~\ref{sec:methods} and present the results in Sec.~\ref{sec:results}. We discuss the implications in Sec.~\ref{sec:discussions} and draw our conclusions in Sec.~\ref{sec:conclusions}


\section{Methods and Model considerations \label{sec:methods}}

\subsection{Atomic and molecular gas distribution\label{subsec:gas_model}}

\begin{figure}
\centering    
    \subfigure[Molecular Hydrogen]{
    \includegraphics[width=0.48\textwidth]{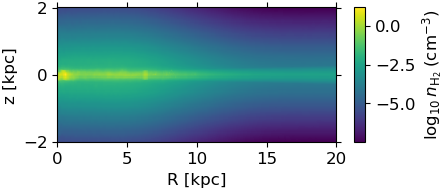}}
    \hfill
    \subfigure[Atomic Hydrogen]{
    \includegraphics[width=0.46\textwidth]{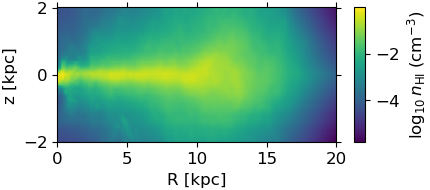}}
    \caption{Axisymmetric density distribution of (a) molecular H$_2$ and (b) atomic 
H\,{\sc i} in Galactocentric coordinates, obtained by azimuthally
averaging the 3D interstellar medium reconstruction of
\citet{2025A&A...693A.139S}.}
    \label{fig:h2_dist}
\end{figure}

Galactic molecular clouds dominate the H$_2$ distribution and play an
important role in shaping the observed CR spectrum
\citep{Biswas:2018gxd,Biswas:2019vkq, DeSarkar:2019tjy}. The central molecular zone
(CMZ), extending to $\sim200$ pc from the Galactic center, contains a high
molecular gas density
\citep{2007A&A...467..611F,2016ApJ...822...52R,2020MNRAS.493..351C,
2025ApJS..280...16W}. Earlier Galactic gas models often used factorized
density profiles, such as the prescription of \citet{Lipari:2018gzn},
based on the H\,{\sc i} measurements of \citet{Kalberla:2008uu} outside
the Galactic bulge.

We use the H\,{\sc i} and H$_2$ density cubes from the 3-D ISM reconstruction of \citet{2025A&A...693A.139S},
which provides a spatially coherent model of the full gas distribution. The Sun-centered Cartesian grids are transformed to Galactocentric cylindrical coordinates and azimuthally averaged to obtain
the axisymmetric gas distributions $n_{\rm H\textsc i}$ and $n_{\mathrm{H}_2}$ as a function of $(R,z)$.
The
resulting gas maps are shown in Fig.~\ref{fig:h2_dist}. The total
hydrogen density is $n_{\rm H}=n_{\rm H\textsc i}+2n_{\mathrm{H}_2}$.
The ISM helium abundance is $n_{\rm He}/n_{\rm H}\sim0.1$ and hence is not
included in our analysis.


\subsection{Interstellar radiation field and optical depth\label{subsec:isrf}}

At very high energies (VHE; $E_\gamma \gtrsim 0.1~{\rm TeV}$),
Galactic $\gamma$-rays are attenuated by pair production on ambient
radiation fields, including starlight, dust-reprocessed IR photons,
and the CMB. Absorption due to the far-IR
component becomes important above a few tens of TeV, with the optical depth
typically peaking around $E_\gamma\simeq100$--$200~{\rm TeV}$, whereas CMB
absorption dominates at $E_\gamma\gtrsim300~{\rm TeV}$ \citep{Vernetto:2016alq}. At $E_\gamma\geq1~{\rm TeV}$, attenuation by starlight is
subdominant compared to that by IR photons and the CMB. The
extragalactic background light can also be neglected for Galactic
propagation, since its effect is relevant mainly over extragalactic path
lengths. The CMB is treated as an isotropic blackbody with
$T_{\rm CMB}=2.7255~{\rm K}$ \citep{Planck:2019evm}, and its redshift evolution is negligible over
Galactic distances.

For a $\gamma$-ray arriving from direction $(l,b)$
and line-of-sight distance $s$, the IR optical depth $\tau_{\gamma\gamma}^{\rm IR}(E_\gamma,l,b,s)$ is obtained by integrating
the inverse mean free path $\lambda_{\gamma\gamma}^{-1}(E_\gamma,l,b,s^\prime)$
over the line-of-sight variable $s^\prime$.
The mean free path for the anisotropic IR photon field is,
\begin{align}
\lambda_{\gamma\gamma}^{-1}
=
\int d\Omega \int d\epsilon\,
\frac{dn_{\rm IR}}{d\epsilon\,d\Omega}
(\epsilon,\Omega;\mathbf{x})
(1-\cos\theta)\,
\sigma_{\gamma\gamma}(E_\gamma,\epsilon,\theta),
\end{align}
where $dn_{\rm IR}/d\epsilon d\Omega$ is the IR photon number density at
energy $\epsilon$, direction $\Omega$, and position $\mathbf{x}$.
$\sigma_{\gamma\gamma}$ is the pair-production cross section
\citep{Gould_1967}, and $\theta$ is the angle between the
$\gamma$-ray and target photon.

The IR emission from interstellar dust is modeled as a modified blackbody spectrum, with emissivity
\begin{align}
\eta_\lambda
\simeq
\rho_{\rm dust}(R,z)\,
\kappa_\lambda\,
B_\lambda[T(R,z)] ,
\end{align}
where $\rho_{\rm dust}$ is the dust mass density, $\kappa_\lambda$ is the
wavelength-dependent dust emissivity cross section \citep{Weingartner_2001}, and $B_\lambda$ is the
blackbody spectrum at the local dust temperature. At each point along the $\gamma$-ray path, $\eta_\lambda$ is integrated
along each incoming ray direction $\hat\Omega$ to obtain
$I_\lambda(\Omega,\mathbf{x})$, which in turn gives
$dn_{\rm IR}/d\epsilon d\Omega$. 
The optical depth depends on $E_\gamma$, $(l,b,d)$, the dust density, temperature, $\kappa_\lambda$, and the angular
distribution of the IR field. 

For the baseline IR model, we adopt the Galactic dust distribution of
\citet{Misiriotis:2006qq}, with cold and warm dust components modeled as
exponential disks,
\begin{equation}
\rho_{c,w}(R,z)
=
\rho_{c,w}^{0}
\exp\left(
-\frac{R}{R_{c,w}}
-\frac{|z|}{Z_{c,w}}
\right).
\label{eqn:dust_density}
\end{equation}
Here $\rho_{c,w}^{0}$ are the central dust mass densities, and $R_{c,w}$ and $Z_{c,w}$ are the radial and vertical scales of cold and warm dust, respectively. In the baseline model, we have $\rho_c^0=1.51\times10^{-25}~{\rm g~cm^{-3}}$ and $\rho_w^0=1.22\times10^{-27}~{\rm g~cm^{-3}}$, so the cold-dust density
normalization is much larger than that of the warm component.

This model provides a useful large-scale description of the Galactic dust
distribution based on COBE/DIRBE and FIRAS data
\citep{1992ApJ...397..420B,1998ApJ...508...25H,Kelsall:1998bq,Fixsen:1996nj}.
However, the inner-Galaxy dust density is inferred by fitting a
parametric dust distribution to line-of-sight far-IR emission, and the central
$10^\circ\times10^\circ$ region was masked in the DIRBE maps used for the
spatial fit. We therefore adopt this profile as our baseline and vary its inner radial dependence, as summarized in Table~\ref{tab:dust_evolution},
to estimate the systematic uncertainty in the IR optical depth. Each profile is normalized to match the baseline model value at the solar
neighborhood. These variations increase dust density at the Galactic center by factors of a few to nearly two orders of magnitude in the extreme M4 case. 
We calculate $\tau_{\gamma\gamma}(E_\gamma,\ell,b,s)$ on a grid with
$\ell\in[15^\circ,235^\circ]$ in $2^\circ$ steps and $b\in[-5^\circ,5^\circ]$ in $0.5^\circ$ steps (since $\sin b\approx b$ for $|b|\ll1$). Along each line of sight, $\tau_{\gamma\gamma}$ is calculated 
up to the distance at which the Galactocentric radius reaches $R=15~{\rm kpc}$.
The LHAASO diffuse $\gamma$-ray measurements then provide a natural testbed
for assessing how uncertainties in the inner-Galaxy IR field affect the
interpreted Galactic emission at $\gtrsim100~{\rm TeV}$.


\begin{table}
\centering
\caption{
Multiplicative normalization factors $A_{c,w}$ applied to $\rho_{c,w}^0$ in
Eq.~\ref{eqn:dust_density} for different radial profiles. The factors are
chosen so that the dust density, and hence the IR emissivity, matches the
baseline model value at $R_\odot=8.5~{\rm kpc}$, in the solar neighborhood. The default scale lengths are $R_c=5~{\rm kpc}$ and
$R_w=3.3~{\rm kpc}$ for cold and warm dust, respectively.
}
\begin{tabular}{llcc}
\hline
Model & Radial profile & $A_c\,\rho_c^0$ & $A_w\,\rho_w^0$ \\
\hline
Base & $A_{c,w}\,\rho^0_{c,w}\exp(-R/R_{c,w})$ & $\rho_c^0$  & $\rho_w^0$ \\
M1 & $A_{c,w}\,\rho^0_{c,w}\exp(-2R/R_{c,w})$ & $5.47\,\rho_c^0$  & $13.14\,\rho_w^0$ \\
M2 & $A_{c,w}\,\rho^0_{c,w}\exp[-(R/R_{c,w})^2]$ & $3.29\,\rho_c^0$  & $57.90\,\rho_w^0$ \\
M3 & $A_{c,w}\,\rho^0_{c,w}\exp(-2.5R/R_{c,w})$ & $12.81\,\rho_c^0$ & $47.64\,\rho_w^0$ \\
M4 & $A_{c,w}\,\rho^0_{c,w}\exp(-3.5R/R_{c,w})$ & $70.11\,\rho_c^0$ & $625.9\,\rho_w^0$ \\
\hline
\end{tabular}
\label{tab:dust_evolution}
\end{table}

\subsection{CR injection from SNR distribution}

Supernova remnants (SNRs) are widely considered the main sources of
Galactic CRs through diffusive shock acceleration, with efficient
acceleration supported by magnetic-field amplification and shock geometry
effects \citep[see, e.g.,][]{Xu:2017wsm, Xu:2021clv}. Observations of Tycho's SNR indicate
proton acceleration up to $\gtrsim500~{\rm TeV}$ \citep{2012A&A...538A..81M}.
We adopt $E_{p,\max}=10~{\rm PeV}$, which lies between the
$5$ and $50~{\rm PeV}$ source cutoffs commonly used in KRA$_\gamma$
Galactic diffuse $\gamma$-ray and neutrino models \citep{Gaggero:2014xla, Gaggero:2015xza}. Since nuclear spallation is
negligible at these energies, we restrict the transport calculation to CR
protons.
For a distributed SNR population, the injection in units of number density per energy and time interval is
\begin{align}
Q_p(E,R,z)\propto\,f_{\rm S}(R,z)\left({E}/{E_0}\right)^{-\alpha_p},
\end{align}
where we choose a reference energy $E_0=4$ GeV. We adopt a source distribution that follows the SNR radial profile of \citet{1996A&AS..120C.437C}
multiplied by an exponential vertical dependence \citep{2012JCAP...01..010B},
\begin{align}
f_{\rm S}(R,z)\propto
\left({R}/{R_\odot}\right)^{p}
e^{{{-q({R-R_\odot})}/{R_\odot}}}
e^{-{|z|}/{z_g}}
\label{eqn:snr_dist}
\end{align}
with $R_\odot=8.5$ kpc, $p=1.69$, $q=3.33$, and $z_g=0.2$ kpc.  Thus, our injection spectrum reduces to a function of $E$, $R$, and $z$. 
At injection, the spectra required in
Galactic CR models are typically steeper than the canonical first-order
Fermi prediction, with $\alpha_p\simeq2.2$--$2.4$
\citep{Caprioli_2012}. We adopt a conservative choice $\alpha_p=2.4$.

\subsection{CR diffusion and transport}


Assuming cylindrical symmetry and an azimuthal large-scale magnetic field,
$\mathbf{B}_0=B_\phi(R,z)\hat{\phi}$, transport in the $R$--$z$ plane is
perpendicular to the field. We adopt the spatially dependent diffusion
coefficient of \citet{DeLaTorreLuque:2025zsv},
\begin{align}
D(E,R)
=
D_0 \,\beta\,
\left({E}/{E_0}\right)^{\delta(R)}
\end{align}
where $\delta(R)=0.04\cdot(R/\text{1\,kpc}) + 0.17$ for $R<R_\odot$ and $\delta(R)=\delta(R_\odot)$ for $R\geq R_\odot$.
The diffusion coefficient at $E_0=4$ GeV is $D_0=6.1\times 10^{28}$ cm$^2$ s$^{-1}$. 
In Galactocentric $(R,z)$ coordinates, the transport equation including
diffusion and $pp$ losses, and neglecting ionization and Coulomb losses
above $1$ TeV, is \citep{1964ocr..book.....G, Strong:1998pw}
\begin{align}
\frac{1}{R}\frac{\partial}{\partial R}
\left(
R\,D\frac{\partial N_p}{\partial R}
\right)
+
\frac{\partial}{\partial z}
\left(
D\frac{\partial N_p}{\partial z}
\right) & \nonumber \\
-\frac{N_p}{\tau_{pp}}
+Q_p(E,R,z)
&=\frac{\partial N_p}{\partial t}\, ,
\label{eqn:diff}
\end{align}
where $Q_p$ is the source term, and $\tau_{pp}^{-1}= cn_H\sigma_{pp}^{\rm inel}$ is the $pp$ collision rate, $\sigma_{pp}^{\rm inel}$ is the inelastic $pp$ interaction cross-section \citep{Kelner:2006tc} and $N_p (E, R, z)=dN/dE_p$. We also note that convection and re-acceleration are not important for the energy range of our study. In addition, the proton secondary component becomes important only at energies below $\sim 10$ GeV \citep[see, e.g.,][]{Evoli:2017vim}. 

%
    

We choose $\alpha_p$ such that the propagated spectral index, approximately $\alpha_p+\delta(R)$, matches the observed CR proton spectrum in the solar neighborhood. In the steady state, $\partial N_p/\partial t=0$. We solve Eq.~\ref{eqn:diff} numerically in the steady state using a finite difference method over the galactocentric radius range $0<R<15.0~{\rm kpc}$ and $|z|<2.0~{\rm kpc}$. We impose free-escape boundary conditions,
$N_p(E,R_{\rm max},z)=N_p(E,R,\pm z_{\rm max})=0$, and regularity at the
symmetry axis, $\partial N_p/\partial R=0$ at $R=0$.



\section{Results\label{sec:results}}

The LHAASO WCDA and KM2A measurements provide the diffuse Galactic
$\gamma$-ray spectrum in the inner
$(15^\circ<l<125^\circ,\ |b|<5^\circ)$ and outer
$(125^\circ<l<235^\circ,\ |b|<5^\circ)$ Galactic-plane regions
\citep{LHAASO:2023gne,LHAASO:2024lnz}. The inner-region spectrum shows a
softening above a few tens of TeV, with a change in photon index of
approximately $0.5$, while the evidence for a break in the outer region is
weaker. 


\subsection{Leptonic emission from Pulsar Wind Nebulae\label{subsec:leptonic}}

\begin{figure}
    \centering
    \includegraphics[width=0.48\textwidth]{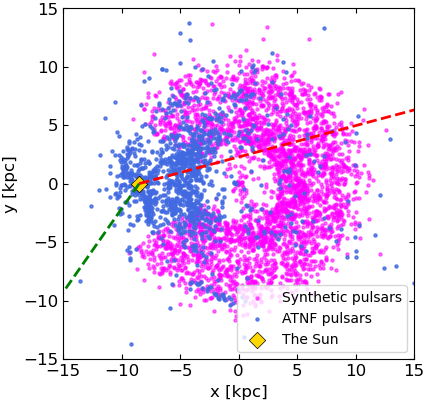}
    \caption{Galactic pulsar distribution, generated from the
observed ATNF pulsar distribution in a $60^\circ$ Galactocentric azimuthal
sector centered on the Sun--Galactic-center line. The distribution
corresponds to one random realization with an aligned-pulsar fraction of
unity. The red dashed line marks $\ell=15^\circ$; the LHAASO region of interest extends anticlockwise to the green dashed line at $\ell=235^\circ$.}
    \label{fig:pulsar_dist}
\end{figure}

\begin{figure*}
    \centering
    \includegraphics[width=0.98\textwidth]{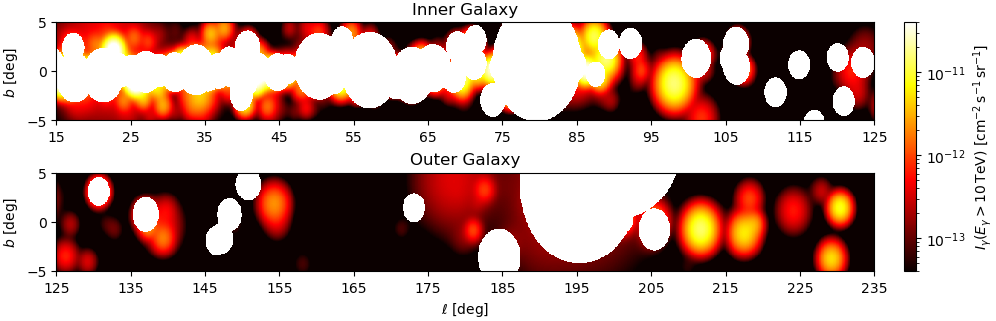}
    \caption{Diffuse integrated $\gamma$-ray emission at $>10$ TeV in the Galactic plane from PWNe for one random realization of the Galaxy. The emission from the white regions are masked out to match the LHAASO analysis \citep{LHAASO:2023gne}. The upper and lower panels correspond to the LHAASO inner ($15^\circ<\ell<125^\circ$, $|b|<5^\circ$) and outer ($125^\circ<\ell< 235^\circ$, $|b|<5^\circ$) regions.}
    \label{fig:pulsar_sky_map}
\end{figure*}

We estimate the unresolved contribution of PWNe and pulsar halos to the diffuse Galactic $\gamma$-ray emission by generating a synthetic population of their parent pulsars.
We use the ATNF\footnote{https://www.atnf.csiro.au/research/pulsar/psrcat/}
pulsar catalog \citep{Manchester:2004bp} to estimate the age ($\tau_{\rm age}$) and radial
distributions of pulsars in a $60^\circ$ Galactocentric azimuthal sector
that contains the Sun. We impose $|z|\leq1.7~{\rm kpc}$ corresponding to the LHAASO vertical extent and
$\tau_{\rm age}<10~{\rm Myr}$ to exclude older pulsars.
After applying an age-dependent beaming fraction $\epsilon_{\rm b}$
\citep{TM98}, the 737 pulsars selected in the reliable sector correspond
to a total population of $N_{\rm true}=737/\epsilon_{\rm b}$, accounting
for misaligned pulsars.

We first generate a Monte Carlo realization of the reliable sector using
the inferred age and radial distributions, including Gaussian fluctuations of width $\sqrt{N_{\rm true}}$. The same procedure is then
repeated over the remaining Galactocentric azimuthal sectors to construct
the full Galactic population. An example random realization is shown in Fig.~\ref{fig:pulsar_dist}
for $\epsilon_{\rm b}=1$. The dashed red and green lines indicate the LHAASO region of interest in the Galactic plane. For each source, we convert the pulsar age to the spin-down power $\dot{E}$ using $\log_{10}(\dot{E}/{\rm erg~s^{-1}}) = -1.529\log_{10}(\tau_{\rm age}/{\rm yr})+42.56$, and estimate the
maximum $\gamma$-ray energy of the associated PWNe as $E_{\gamma,\max}=0.9(\dot{E}/10^{36}\,{\rm erg~s^{-1}})^{0.65}~{\rm PeV}$
following \citet{WilhelmideOna:2022zmp}.

The ATNF catalog is not a complete census of Galactic pulsars, so we use the observed pulsars only as empirical templates for the age and radial distributions of the parent population. The cut $\tau_{\rm age}<10$ Myr removes old pulsars with low present spin-down powers, which are unlikely to contribute appreciably to TeV--PeV PWN emission, while $|z|\leq1.7$ kpc matches the vertical extent of the LHAASO field of view. Catalog incompleteness and sampling biases, including radio-selection effects and distance uncertainties, can affect the inferred age and radial distributions. In particular, missing distant inner-Galaxy pulsars or young high-$\dot E$ systems would lead to an underestimate of the unresolved PWN contribution, whereas incompleteness among old low-$\dot E$ pulsars has little impact on the TeV--PeV flux.

The $\gamma$-ray spectrum,
$\Phi_\gamma=N_0(E/E_0)^{-\Gamma}$ is assumed for each PWN, with $E_0=3~{\rm TeV}$ for the LHAASO
WCDA band and $E_0=50~{\rm TeV}$ for the KM2A band \citep{LHAASO:2023rpg}.
Among the 34 LHAASO sources associated with pulsars/PWNe, the spectral
index is available for 31 sources in the KM2A band and 28 sources in the
WCDA band. For each synthetic source, we assign a randomly generated $\Gamma$ within the measured range in the corresponding detector band. For the KM2A and WCDA components, we adopt the fitted relations between the
integrated $\gamma$-ray flux and the pulsar spin-down power derived by
\citet{Kaci:2025myt}, including the same Gaussian scatter. 
For the KM2A component,
\begin{align}
\log_{10}\left(4\pi d^2 F_{>10\,{\rm TeV}}\right)
=
0.942\,\log_{10}(\dot{E}) - 3.894 .
\end{align}
This gives the integrated flux $F_{>10\,{\rm TeV}}$, which is then used to
fix the normalization of the KM2A spectrum up to $E_{\gamma,\rm max}$. If $E_{\gamma,\rm max}\ge 25~{\rm TeV}$, the WCDA normalization is set by requiring
continuity with the KM2A spectrum at $25~{\rm TeV}$. If
$E_{\gamma,\rm max}<25~{\rm TeV}$, only the WCDA component is included, and its
normalization is obtained from the analogous relation
\begin{align}
\log_{10}\left(4\pi d^2 F_{>1\,{\rm TeV}}\right)
=
1.029\,\log_{10}(\dot{E}) - 5.682 ,
\end{align}
using the integrated flux between $1~{\rm TeV}$ and $E_{\max}$. The resulting normalizations are used to compute the spectra in the
WCDA $(1$--$25~{\rm TeV})$ and KM2A $(25$--$1000~{\rm TeV})$ bands.
We assume a physical extension of $20~{\rm pc}$ for each PWN \citep[see, e.g.,][]{HAWC:2017kbo, Martin:2022hrx}. The source flux is then distributed over neighboring sky pixels using a Gaussian kernel that combines the intrinsic angular
extension with the LHAASO point-spread function (PSF). For the PSF, we
adopt the largest value in the $10$--$15~{\rm TeV}$ bin \citep[][see Fig. 7 therein]{Aharonian:2020iou}. Only sources within the LHAASO field of
view are retained.

To identify unresolved sources, we compare the differential flux
normalizations of the WCDA and KM2A components with the corresponding
declination-dependent LHAASO sensitivity curves \citep{LHAASO:2023rpg}. We adopt the most
optimistic detection thresholds, corresponding to a point source with an
$E^{-2}$ spectrum for the WCDA band and an $E^{-3}$ spectrum for the KM2A band. This choice
provides a conservative lower limit on the unresolved PWNe contribution.
For comparison with the diffuse Galactic-plane emission, we apply a source
mask based on detected LHAASO-KM2A sources and known TeVCat sources \citep{LHAASO:2023gne}. The
resulting sky-intensity distribution for one random realization of the
Galaxy is shown in Fig.~\ref{fig:pulsar_sky_map}.

The $\gamma$-ray flux is obtained by summing the contributions of
unresolved sources over the sky pixels, after applying
$\gamma\gamma$ attenuation in the ISRF. The mean unresolved flux,
computed from 1000 random realizations of the Galaxy, is shown in
Fig.~\ref{fig:gamma_model}. Our estimate is consistent with the leptonic flux found by \citet{Kaci:2025myt}. 
For the inner region in our baseline IR model, 
the leptonic
contribution accounts for about $20\%$ of the diffuse flux at a few TeV,
increases to $\sim30\%$ at a few tens of TeV, and decreases to
$\sim10\%$ at the highest energies.


\subsection{Diffuse $\gamma$-ray and neutrino fluxes}

\begin{figure}
    \centering
    \includegraphics[width=0.46\textwidth]{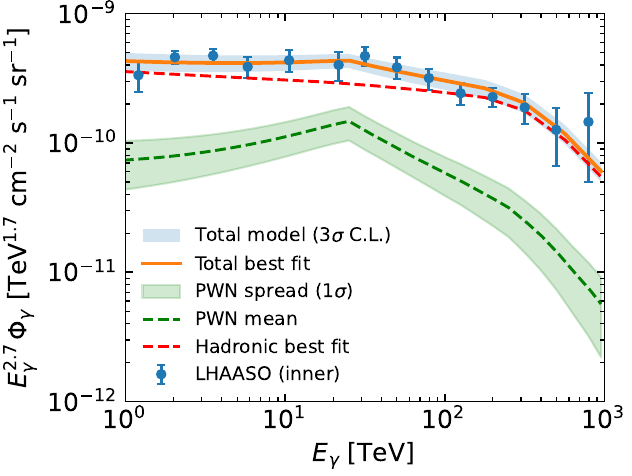}
    \caption{LHAASO inner-region ($15^\circ<\ell<125^\circ$, $|b|<5^\circ$) diffuse $\gamma$-ray spectrum for the baseline model. The blue band shows the $3\sigma$ confidence interval from the fitted hadronic normalization, while the green band shows the $1\sigma$ spread of the unresolved PWN component.}
    \label{fig:m0_confidence}
\end{figure}

\begin{figure*}
    \centering
    \includegraphics[width=0.46\textwidth]{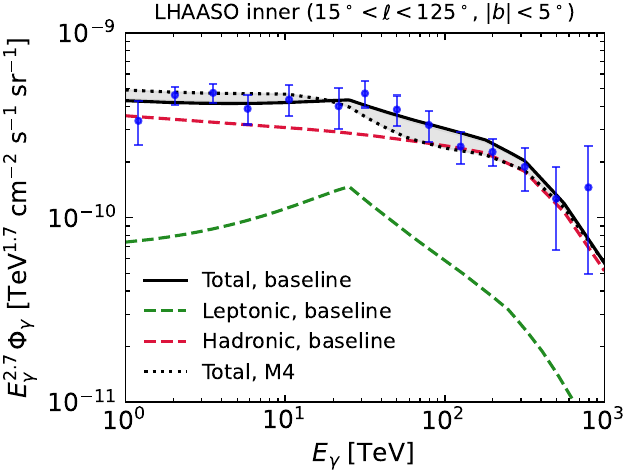}%
    \hspace{0.5cm}
    \includegraphics[width=0.46\textwidth]{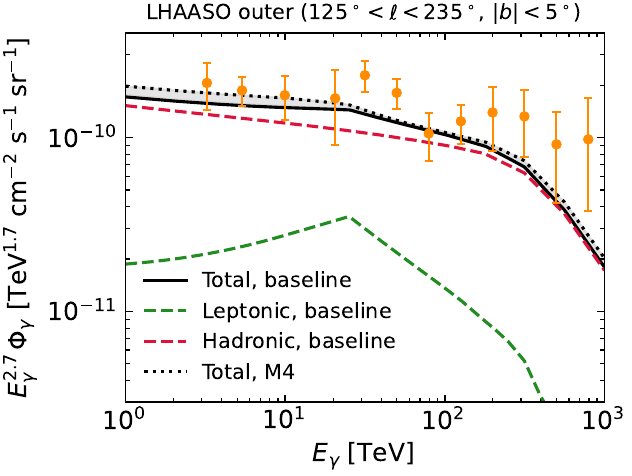}
    \caption{Diffuse $\gamma$-ray flux for the LHAASO inner ($15^\circ<\ell<125^\circ$, $|b|<5^\circ$) and outer ($125^\circ<\ell< 235^\circ$, $|b|<5^\circ$) region of interest. The shaded region indicates the variation due to uncertainties in the IR model stemming from alternate dust density profiles assumed in Table~\ref{tab:dust_evolution}. The black solid line shows the base model while the black dotted line shows the most extreme M4 model. The LHAASO data are shown as blue and orange points for the inner and outer regions, respectively \citep{LHAASO:2024lnz}. 
    }
    \label{fig:gamma_model}
\end{figure*}

\begin{figure*}
    \centering
    \includegraphics[width=0.46\textwidth]{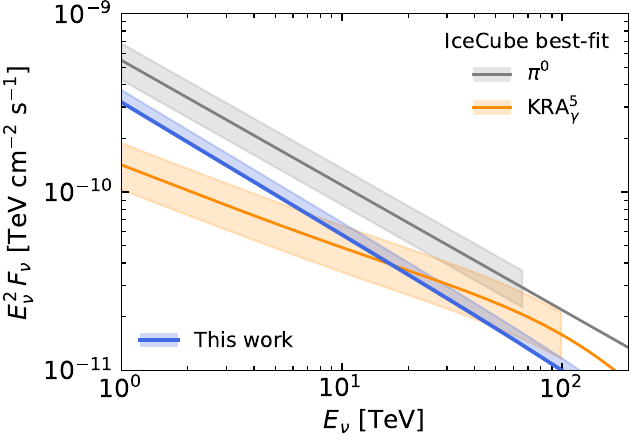}
    \hspace{0.5cm}
    \includegraphics[width=0.46\textwidth]{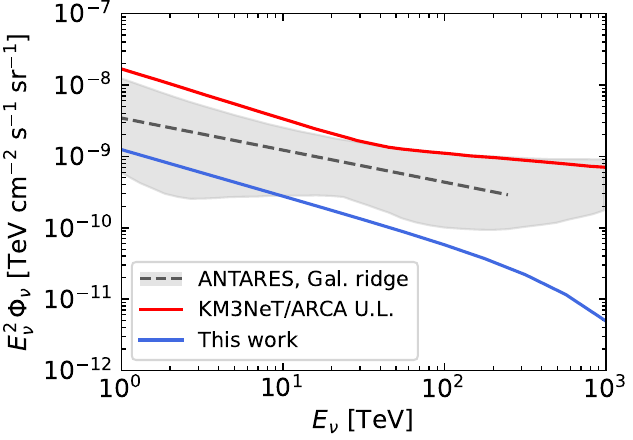}
    \caption{\textit{Left:} Neutrino flux obtained in our baseline model (blue line) and other IR distribution profiles (blue shaded), using the best-fit normalizations obtained in Fig.~\ref{fig:gamma_model}. The IceCube all-sky measurement for two of their best-fit models is shown for comparison \citep{IceCube:2023ame}. \textit{Right:} Neutrino flux from the Galactic ridge ($|\ell|<30^\circ$, $|b|<2^\circ$) in our model compared to the 68\% C.L. upper limit from ANTARES observation \citep{ANTARES:2022izu} and 90\% credible-interval upper limit from KM3NeT using 640 days of data \citep{Filippini:2023bbc}.}
    \label{fig:nu_model}
\end{figure*}

The hadronic $\gamma$-ray emission depends on the gas density, SNR distribution, and CR spectrum.
We use the steady-state CR distribution,
$N_p(E_p,R,z) = dN/dE_p$, to calculate the $\gamma$-ray and neutrino fluxes. The line-of-sight integrated $\gamma$-ray intensity is
\begin{align}
I_\gamma(E_\gamma,l,b)
=
\int \frac{ds}{4\pi}\,
{Q_\gamma(E_\gamma,l,b,s)}\,
e^{-\tau_{\gamma\gamma}(E_\gamma,l,b,s)},
\label{eqn:gamma_intensity}
\end{align}
where $Q_\gamma(E_\gamma,R,z)$ is the local hadronic emissivity, in units
of ${\rm TeV^{-1}\,cm^{-3}\,s^{-1}}$, calculated using the
$pp$-interaction parametrization of \citet{Kelner:2006tc}.
To compare with the LHAASO data, we apply the same source mask to the
hadronic component, removing bright Galactic-plane sources. The
solid-angle-averaged diffuse $\gamma$-ray intensity over the surviving
pixels is
\begin{equation}
\Phi_\gamma(E)
=
\frac{1}{\Omega_{\rm ROI}}
\int_{\rm ROI}
I_\gamma(E,l,b)\,d\Omega ,
\end{equation}
where $d\Omega=\cos b\,db\,dl$ and $\Omega_{\rm ROI}$ is the unmasked
solid angle of the region of interest (ROI).

We fit the LHAASO inner-region data by $\chi^2$ minimization, keeping the leptonic contribution fixed and varying only the hadronic normalization. We use the inner-region spectrum for the fit because it provides a stronger constraint than the outer region and use the total LHAASO uncertainties, with statistical and systematic errors added in quadrature, in the $\chi^2$ calculation. This yields a conservative goodness-of-fit estimate, resulting in $\chi^2_\nu<1$, primarily due to large uncertainties at the highest energies.
The best-fit for the baseline model gives a reduced $\chi^2_\nu\simeq0.5$ for 13 degrees of freedom, shown in Fig.~\ref{fig:m0_confidence}. The blue-shaded band shows the $3\sigma$ confidence interval on the fitted hadronic normalization, obtained from $\Delta\chi^2=9$ for one fitted parameter. The PWN component is fixed to its mean value in the fit, with its $1\sigma$ Monte Carlo spread shown by the green-shaded region.


We repeat the fit for the alternative IR radial-dependence models in Fig.~\ref{fig:gamma_model}. The resulting range is shown by the gray shaded band. The black dotted curve shows the most extreme variation, M4. The fit improves slightly to $\chi_\nu^2\!=\!0.4$ for the stronger M3 profile, but deteriorates for the M4 profile to $\chi_\nu^2=0.8$. Hence, no statistically significant improvement in the fit to the LHAASO data is found for the radial profiles considered here. The ISRF comparison uses the best-fit normalization for each case. Including the normalization uncertainty may moderately broaden the flux range. The modified profiles differ most strongly in the central region, which lies outside the LHAASO longitude range and is further affected by the Galactic-plane source mask. Within the adopted region of interest and source masks, the LHAASO diffuse measurement is therefore robust against the five different IR-background profile uncertainties considered here.

We apply the global normalization of the injected Galactic CR population obtained from the inner-region fit to the outer region and do not perform a separate $\chi^2$ minimization. The gas distribution, PWN population, and ISRF are still evaluated with their spatial dependences for the assumed models. Only the overall injected CR normalization is kept common. The resulting
outer-region flux is shown in the right panel of Fig.~\ref{fig:gamma_model}. The model underestimates the outer-region flux around the spectral break near $\sim 30$ TeV and at the highest energies, suggesting that the baseline model may miss additional outer-Galaxy contributions or require modifications to the assumed CR source, gas, or transport model. 
For the adopted SNR source distribution from \citet{1996A&AS..120C.437C}, the radial profile peaks at $R_{\rm peak}=(p/q)R_\odot$ (see Eqn.~\ref{eqn:snr_dist}). Source distributions with larger $R_{\rm peak}$ can enhance the outer-region flux. Other possible effects include local CR spectral hardening and residual unresolved or extended source emission. The alternative IR profiles have a negligible effect on the LHAASO outer-region spectrum.

The directional neutrino intensity is calculated over the full sky and
integrated over solid angle,
\begin{equation}
F_\nu(E)
=
\int I_\nu(E,l,b)\,d\Omega ,
\end{equation}
where $I_\nu$ is calculated analogously to Eq.~\ref{eqn:gamma_intensity},
but without any optical-depth attenuation. The all-sky single-flavor $(\nu_\mu+\bar{\nu}_\mu)$ flux, after accounting for neutrino oscillations, is shown in the left panel of Fig.~\ref{fig:nu_model}. The blue solid curve
shows the baseline model. The blue-shaded band gives the range from the
hadronic normalizations used to fit the inner-region $\gamma$-ray data for
the different IR models.
We also show the IceCube constraints on Galactic neutrino
emission derived for the $\pi^0$, KRA$_\gamma^5$, and KRA$_\gamma^{50}$
templates \citep{IceCube:2023ame,Gaggero:2015xza}. The shaded bands show the $1\sigma$ uncertainty of the IceCube best-fit template fluxes over the neutrino-energy range contributing 90\% of the
detection significance. The $\pi^0$ template
extrapolates the GeV $\pi^0$-decay component inferred from $\gamma$-ray data with an $E^{-2.7}$ spectrum and fixed morphology. The
KRA$_\gamma$ templates allow a spatially dependent CR spectrum, producing harder
emission toward the Galactic center. Our prediction is consistent with IceCube estimates and
approaches the KRA$_\gamma^5$ best-fit flux beyond a few hundred TeV.

\subsection{Galactic center region}

\begin{figure*}
    \centering
    \includegraphics[width=0.46\textwidth]{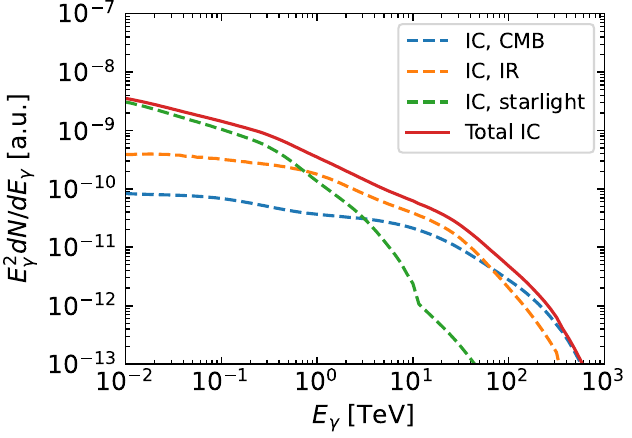}
    \hspace{0.5cm}
    \includegraphics[width=0.46\textwidth]{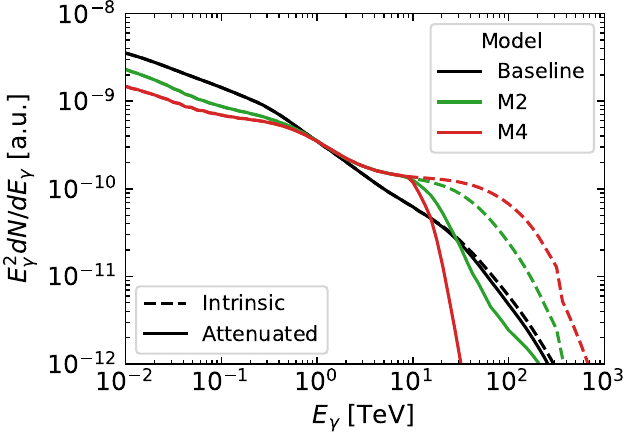}
    \caption{Flux of inverse-Compton $\gamma$-ray emission from a steady-state electron spectrum, shown in arbitrary units. The source is placed near the CMZ at $R=0.2$ kpc. The left panel shows the relative contributions of the different ISRF photon fields in the baseline model. The right panel shows the change in the total spectrum for different IR radial-dependence models.}
    \label{fig:ic_model}
\end{figure*}

Using the hadronic normalization fixed by the LHAASO inner-region fit, we
estimate the neutrino flux from the Galactic Ridge, defined as the region
$|\ell|<30^\circ$ and $|b|<2^\circ$. This region is expected to provide
one of the strongest Galactic neutrino signals because of its high gas
column density and enhanced CR activity \citep{Neronov:2013lza, Yang:2016jda, Liu:2016lzx, Gaggero:2017jts}. The CMZ is an asymmetric layer of predominantly
molecular gas in the central few hundred parsecs of the Galaxy, extending to $\sim200~{\rm pc}$ from the Galactic center. 
Our transport model is axisymmetric in $(R,z)$, so a fully 3-D Galactic-center transport calculation would be needed to
capture the coupled effects of non-axisymmetric gas, source, and
magnetic-field structure \citep{Dorner:2023dqd}.
To account for the enhanced central gas concentration, we add the CMZ
atomic and molecular hydrogen components from
\citet{2007A&A...467..611F} to the gas model described in
Sec.~\ref{subsec:gas_model}. 

The CMZ gas is included only in the
line-of-sight emissivity. This is justified by comparing the diffusion
time, $t_{\rm diff}\simeq L^2/6D(E,R)$, with the hadronic loss time, $t_{pp}\simeq (n_{\rm H}c\sigma^{\rm inel}_{pp})^{-1}$. Taking $L=0.2~{\rm kpc}$ as the characteristic CMZ escape scale and evaluating the diffusion coefficient at a representative radius $R=0.1~{\rm kpc}$, we obtain $t_{\rm diff}\simeq8.5\times10^3~{\rm yr}$ at $E_p=10~{\rm TeV}$. For a representative CMZ density $n_{\rm H}=150~{\rm cm^{-3}}$ and $\sigma_{pp}^{\rm inel}\simeq5\times10^{-26}~{\rm cm^2}$, the hadronic loss time is $t_{pp}\simeq1.4\times10^5~{\rm yr}$. Since $t_{\rm diff}\ll t_{pp}$, the
CMZ gas mainly changes the emissivity rather than the propagated proton density.

We use line-of-sight distance bins of $0.1~{\rm kpc}$ for the Galactic Ridge region, with finer $0.01~{\rm kpc}$ binning for $R<500~{\rm pc}$. The neutrino flux in our baseline model from the Galactic Ridge per unit solid angle is shown by the blue solid line in the right panel of Fig.~\ref{fig:nu_model}. The shaded band shows the 68\% containment region of the ANTARES
Galactic Ridge neutrino flux inferred from the posterior distribution of
the power-law normalization and spectral index, while the dashed line
shows the corresponding best-fit flux \citep{ANTARES:2022izu}. 
We also show the 90\% credible-interval upper limit on the neutrino flux
from the Galactic Ridge obtained with 640 days of KM3NeT/ARCA data \citep{Filippini:2023bbc}.

The Galactic Ridge neutrino comparison serves as a consistency check on the hadronic component inferred from the LHAASO diffuse $\gamma$-ray fit. An overly hard global hadronic component would approach the all-sky IceCube diffuse flux, while present Ridge limits may allow local CR hardening near the Galactic Center relative to our baseline model. Next-generation $\gamma$-ray detectors, such as CTA, will provide an important complementary probe, although a quantitative forecast would require a dedicated sensitivity study \citep{Neronov:2020zhd}.

The axisymmetric IR photon density profiles considered here are centrally peaked, so their effect should be stronger for individual sources near the Galactic center, even though the LHAASO inner-region diffuse flux is affected only modestly. Ultrahigh-energy Galactic $\gamma$-ray sources extending to
$\gtrsim50~{\rm TeV}$ have been observed, with IC emission on CMB photons dominating at the highest energies \citep[see, e.g.,][]{Boxi:2025idb}. However, an enhanced IR photon field can increase the IC emission on the
ISRF while also strengthening $\gamma\gamma$ attenuation through a larger
optical depth. The steady-state electron distribution is
\begin{align}
N_e(E_e,R,z)
=
\frac{1}{b_{\rm tot}}
\int_{E_e}^{E_{e,\max}}
Q_e(E_e',R,z)\,dE_e' ,
\label{eqn:steady_electron}
\end{align}
where $b_{\rm tot}(E_e, R,z) \equiv -dE_e/dt=b_{\rm syn}+b_{\rm IC}$ is the sum of the synchrotron and IC energy-loss rates. For electrons of energy $E_e$ scattering target photons of energy
$\epsilon$, the IC emissivity scales approximately as
$Q_{\gamma,\rm IC}\propto n_{\rm ph}(\epsilon,R,z)/b_{\rm tot}(E_e,R,z)$.
In the synchrotron-dominated regime, an enhanced IR photon density increases the IC flux. In the IC-dominated regime, however, the
increase in $n_{\rm ph}$ is partly offset by stronger cooling through
$b_{\rm tot}$, and the dominant effect of a stronger IR profile is the larger $\gamma\gamma$ attenuation. Hence, the IR radial distribution models are not expected to violate the Fermi-LAT constraints on the diffuse background emission at GeV--TeV energies in the same ROI as LHAASO \citep{Zhang:2023ajh}.

We use secondary electrons from $pp$ interactions as a proxy to study how
centrally enhanced IR photon-density profiles affect the observed
spectrum. The left panel of Fig.~\ref{fig:ic_model} shows the relative contributions of the background photon fields to the total IC spectrum in the baseline model, for a source near the CMZ, at $R\simeq200~{\rm pc}$ along $(\ell,b,d)=(0^\circ,0^\circ,8.3~{\rm kpc})$. Above a few TeV, IC emission on the IR photon field becomes important. We assume an RMS Galactic magnetic field of $3~\mu{\rm G}$, for which synchrotron losses dominate above $E_e\sim100$ TeV due to Klein--Nishina suppression of IC scattering.
The right panel of Fig.~\ref{fig:ic_model} shows the IC spectra for different IR models. More centrally peaked profiles harden the spectrum in the WCDA band, while the extreme M4 case sharply attenuates the flux above a few tens of TeV. Future multiwavelength measurements of point sources near the Galactic center can therefore constrain the synchrotron emissivity and, in turn, the radial IR density distribution in IC-dominated scenarios \citep[see, e.g.,][]{Gupta:2024hxl, Mizuno:2026vaz}. An anisotropic radiation field limits a direct extrapolation of this result to all sources in the Galactic plane \citep[see also][]{DiMarco:2024kgd}.

\section{Discussions and Summary\label{sec:discussions}}

We fit the LHAASO diffuse $\gamma$-ray flux in the inner and outer ROIs
with leptonic PWN and hadronic diffuse-CR components. Although the LHAASO region excludes the Galactic-center direction, inner-Galaxy lines of sight
pass within a few kpc of the center, making the inner dust profile relevant
for the IR optical depth. We use the \citet{Misiriotis:2006qq} profile as
the baseline IR model and vary its inner radial dependence to estimate the systematic uncertainty.

For the adopted source distribution and transport model, the alternative ISRF prescriptions slightly improve the fit to the LHAASO diffuse $\gamma$-ray data, but the preference remains modest. LHAASO has limited leverage in the innermost region because of its longitude range and source masking. 
For the adopted diffuse regions and source masks, the LHAASO diffuse measurement is only weakly sensitive to the IR-background profile variations tested here.

The hadronic component in our model is produced by Galactic CRs propagating through interstellar gas. Source cocoons could provide an additional hard component that is not considered here \citep{Ambrosone:2025wxc,Yang:2024igs}. We normalize the unresolved leptonic component using the empirical flux--spin-down relation of \citet{Kaci:2025myt}, finding a
$\sim\!30\%$ contribution at a few tens of TeV, consistent with their data-driven estimates.

Our outer-ROI fit underpredicts the highest-energy flux, indicating that an additional source component may be required. This outer-region tension may also depend on the assumed radial source distribution. The SNR source distribution adopted here peaks at $\sim4~{\rm kpc}$ \citep{1996A&AS..120C.437C,Case:1998qg}. Source profiles shifted farther outward can enhance the outer-region flux, whereas distributions with lower $p/q$ in Eqn.~\ref{eqn:snr_dist} reduce it \citep[see, e.g.,][]{Green:2015isa}. More recent studies suggest that a purely exponential radial profile may better describe the present SNR sample \citep{Verberne:2021tse,Ranasinghe:2022ntj}.



The CR proton spectral index $\alpha_p=2.4$ is motivated by the observed local proton spectrum, which is roughly $E^{-2.6}$ in the low-TeV range \citep{DAMPE:2019gys}, softens to about
$E^{-2.8}$--$E^{-2.9}$ above $\sim10~{\rm TeV}$
\citep{CALET:2022vro}, hardens near
$166~{\rm TeV}$ \citep{2024PhRvL.132e1002V}, and shows a broad PeV-scale hump \citep{LHAASO:2025byy}. In the diffusion-dominated regime, the propagated slope is approximately $\alpha_p+\delta(R)$.
The resulting $\gamma$-ray slope is further modified by the $\sim E^{0.1}$ increase of the $pp$ cross section above
threshold and by $\gamma\gamma$ attenuation.

Our results should be interpreted within the assumed spatially dependent diffusion-coefficient model.
A nonuniversal diffusion coefficient depends on the properties of the turbulent magnetic fields in the ISM \citep[e.g.,][]{Lazarian:2021kvd}.
Recent analysis has shown that uncertainties in the gas distribution, hadronic cross section, local CR spectrum, and CR spatial profile can significantly reduce the apparent tension between the LHAASO diffuse data and standard hadronic predictions \citep{Vecchiotti:2024kkz}.

The Galactic Center and much of the expected Galactic-plane neutrino emission lie in the Southern sky, where IceCube track searches are strongly affected by atmospheric-muon background. The Galactic-plane measurement therefore uses cascade events,
which reduce atmospheric-neutrino contamination by about an order of magnitude at TeV energies but have poorer angular resolution than tracks
\citep{IceCube:2023ame}. The neutrino flux obtained here is consistent with the IceCube
all-sky flux estimate. However, unresolved Galactic sources may contribute a quasi-diffuse component to the IceCube Galactic plane signal, especially near $100~{\rm TeV}$ \citep{Ambrosone:2023hsz}. 

Current IceCube data do not yet distinguish diffuse neutrinos produced by CR interactions with the interstellar medium from unresolved sources or extended source regions \citep{IceCube:2023ujd}. The Galactic Ridge region, $|l|<30^\circ$ and $|b|<2^\circ$, includes the CMZ and the innermost Galactic plane, and is therefore sensitive to possible variations of the CR spectrum toward the Galactic Center.
With the hadronic normalization fixed by the LHAASO inner-region fit, the predicted Galactic Ridge neutrino flux is about a factor of two below the ANTARES best fit, while current KM3NeT/ARCA data show no significant excess \citep{Filippini:2023bbc}.


\section{Conclusions\label{sec:conclusions}}

Stronger radial evolution of the ISRF can significantly modify the $\gamma$-ray SED of point sources, even though its effect on the LHAASO diffuse fit is modest. For point-like sources, the IC emission is partly self-regulated because an enhanced photon background increases the IC target density but also reduces the steady-state electron density through stronger cooling (see Eq.~\ref{eqn:steady_electron}). Both leptonic and hadronic $\gamma$-ray emission are also attenuated by the enhanced $\gamma\gamma$ opacity. As a result, point-like or extended sources near the Galactic Center may show spectral hardening in the WCDA band, while the most centrally enhanced IR
profiles can produce a sharp cutoff above a few tens of TeV. Future KM3NeT observations of Galactic point-like and extended neutrino sources will help isolate the hadronic component. Together with multiwavelength constraints on the synchrotron spectrum and $\gamma$-ray SED measurements of individual sources, this will provide a multimessenger probe of the inner-Galaxy CR population and constrain the radial distribution of the ISRF near the Galactic Center.

\begin{acknowledgments}
S.X. acknowledges the support from the NASA ATP award 80NSSC24K0896 and the NASA Heliophysics Living with A Star Science Program 80NSSC25K0067. 
S.D. thanks Ke Fang and Kohta Murase for useful discussions.
Numerical computations in this work were carried out at the UFIT research computing facility.
\end{acknowledgments}

\bibliography{sample631}{}
\bibliographystyle{aasjournal}



\end{document}